\documentclass[%
 aip,
 amsmath,amssymb,
 reprint,%
]{revtex4-1}

\usepackage{graphicx}
\usepackage{dcolumn}
\usepackage{bm}

\usepackage[utf8]{inputenc}
\usepackage[T1]{fontenc}
\usepackage{mathptmx}

\begin{document}

\preprint{AIP/123-QED}

\title[Model of graphene nanobubble: combining classical density functional and elasticity theories]{Model of graphene nanobubble: \\ combining classical density functional and elasticity theories}

\author{T.F. Aslyamov }
\affiliation{Center for Design, Manufacturing and Materials, Skolkovo Institute of Science and Technology,
  Bolshoy Boulevard 30, bld. 1, Moscow, 121205, Russia}

\author{E. S. Iakovlev}
\affiliation{Center for Design, Manufacturing and Materials, Skolkovo Institute of Science and Technology,
  Bolshoy Boulevard 30, bld. 1, Moscow, 121205, Russia}

\author{I. Sh. Akhatov}
\affiliation{Center for Design, Manufacturing and Materials, Skolkovo Institute of Science and Technology,
  Bolshoy Boulevard 30, bld. 1, Moscow, 121205, Russia}

\author{P. A. Zhilyaev}
\email{Author  to  whom  correspondence  should  be  addressed;  electronic  mail: p.zhilyaev@skoltech.ru}
\affiliation{Center for Design, Manufacturing and Materials, Skolkovo Institute of Science and Technology,
  Bolshoy Boulevard 30, bld. 1, Moscow, 121205, Russia}

\date{\today}

\begin{abstract}
A graphene nanobubble consists of a graphene sheet, an atomically flat substrate and a substance enclosed between them. Unlike conventional confinement with rigid walls and a fixed volume, the graphene nanobubble has one stretchable wall, which is the graphene sheet, and its volume can be adjusted by changing the shape. In this study, we developed a model of a graphene nanobubble based on classical density functional theory and the elastic theory of membranes. The proposed model takes into account the inhomogeneity of the enclosed substance, the nonrigidity of the wall and the alternating volume. As an example application, we utilize the developed model to investigate fluid argon inside graphene nanobubbles at room temperature. We observed a constant height-to-radius ratio over the whole range of radii considered, which is in agreement with the results from experiments and molecular dynamics simulations. The developed model provides a theoretical tool to study both the inner structure of the confined substance and the shape of the graphene nanobubble. The model can be easily extended to other types of nonrigid confinement.
\end{abstract}

\maketitle
There is much experimental evidence of the occurrence of exotic metamorphoses of trapped substances under nanoscopic confinement: square ice emerging inside graphene nanocapillaries~\cite{algara2015square}, the structuring of molecules inside carbon nanotubes and nanopores~\cite{koga2001formation, fomin2015behavior, fomin2015cyclohexane}, the considerable increase in the dielectric constant of water trapped between a diamond surface and graphene~\cite{lim2013hydrothermal}, gases ordering into a crystalline array inside nanopores~\cite{vom1984pressure}, the transition of simple liquids to the solid state caused by volume limitation~\cite{klein1995confinement} and many others. In contrast, in some cases, the confinement itself can be highly influenced by the enclosed substance \cite{gor2018elastocapillarity}. For example, it is known that adsorbed molecules in nanopores induce considerable stresses of approximately several GPa~\cite{long2011pressure}. These stresses result in deformation~\cite{ravikovitch2006density} that can lead to dramatic changes in the confinement structure~\cite{coudert2016adsorption}. Therefore, in a general case, one has to consider a mutual influence of the confinement and the trapped substance in a self-consistent manner.

One appealing example in which an enclosed substance significantly affects confinement and vice versa is a graphene nanobubble (GNB). It consists of a graphene sheet attached to an atomically flat substrate with a trapped substance between them. Originally, GNBs were treated as manufacturing defects during the assembly of van der Waals (vdW) heterostructures (different types of 2D crystals stacked together). However, subsequently, many intriguing and special features of GNBs have been discovered. For instance, the outer graphene sheet of GNBs under strain creates gigantic pseudomagnetic fields~\cite{levy2010strain, qi2014pseudomagnetic}. GNBs can be utilized as a container to visualize chemical reactions~\cite{mu2012visualizing}. In addition, GNBs are spots of intense photoluminescence emission caused by strained-induced variations in the band structure~\cite{tyurnina2019strained}.

Of particular interest is the structure of the substance inside a GNB and its connection with the shape of the bubble. By probing the shape of the GNB, one can implicitly determine the phase state of trapped matter and adhesion energies~\cite{khestanova2016universal, khestanova2018van, iakovlev2019obtaining}. The structure of matter inside GNBs ranges from crystal clusters for GNBs with radii on the order of nanometers~\cite{zamborlini2015nanobubbles, larciprete2016self} to incompressible fluids for GNBs with radii larger than 100~nm~\cite{khestanova2016universal, sanchez2018mechanics}. The shape of GNBs is closely related to the structure and properties of the substance trapped inside them. For spherical GNBs, the phenomenon of the "universal shape" (constant height-to-radius ratio) was experimentally found~\cite{khestanova2016universal}. Subsequent theoretical studies~\cite{khestanova2018van, iakovlev2019obtaining} led to the conclusion that the "universal shape" is a consequence of the constant adhesion energies in the considered range of GNB radii. Another intriguing phenomenon is the existence of exotic 'pancake' GNBs~\cite{khestanova2018van, iakovlev2017atomistic}, which have flat forms with low height-to-radius ratios. Presumably, molecules or atoms of the substance trapped in such bubbles are highly ordered and arranged in a layered structure~\cite{iakovlev2017atomistic}.

\begin{figure}
\includegraphics[width=0.9\columnwidth]{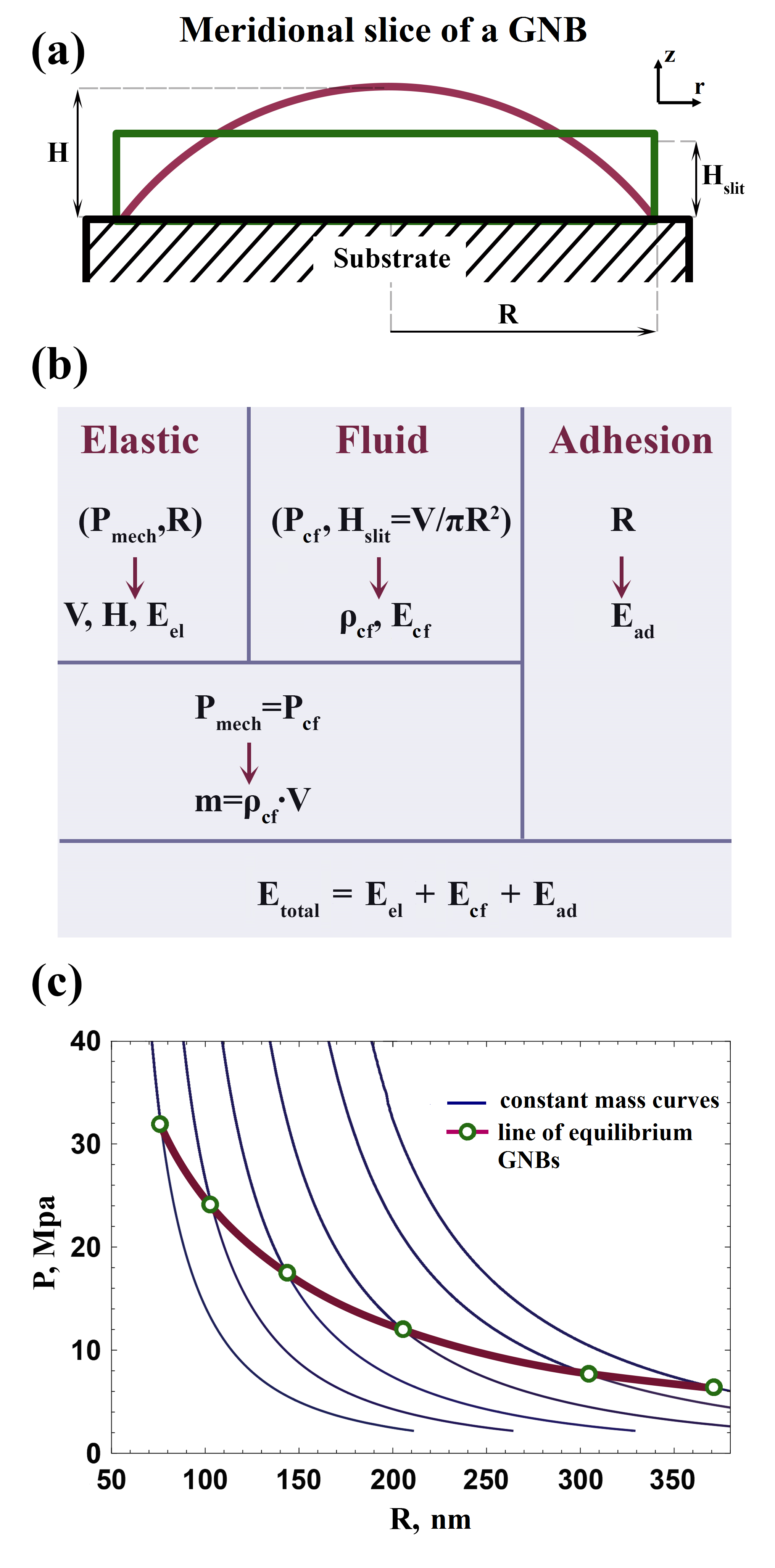}
\caption{\label{fig:fig1} (a) Schematic representation of a GNB. The height $H$ of a GNB is the distance between the top point of the bubble and the surface of the undisturbed graphene sheet. The rectangle with a height $ H_{slit}$ represents the transformed geometry that is used in DFT calculations; (b) visual diagram of the model, which describes the algorithm of the total energy calculation; (c) constant mass curves and a line of equilibrium GNBs filled with argon in ($P$, $R$) coordinates. Constant mass curves are calculated for a fixed mass and specified range of radii, and then an equilibrium GNB with minimum energy is located.}
\end{figure}

Although a number of experimental studies have shown that various condensation phases can exist inside GNBs and that their structure and phase state are mainly determined by the radius of the bubble, there are still many unresolved questions and a great demand for advanced theoretical models that could provide more insights into the structure of the substance enclosed inside GNBs and its connections with the bubble's shape. To address this issue, we develop a GNB model based on classical density function theory (DFT) and the elastic theory of membranes. Classical DFT is applied to obtain the density distribution of the confined substance and evaluate its Helmholtz free energy~($E_{cf}$). The confined substance is assumed to have a noncrystalline structure, and in the following discussion, we will refer to it as the "confined fluid". The mechanical properties of the membrane (graphene sheet) are described by the conventional theory of elasticity~\cite{landau1959course}. This theory is used to calculate the elastic energy ($E_{el}$) of the graphene sheet and determine the height profile of the bubble and the distribution of in-plane deformations. In the model, the footprint radius of the bubble $R$ (see Fig.~\ref{fig:fig1}a) may vary, and the change in the corresponding adhesion energy ($E_{ad}$) is calculated as:

\begin{equation}
    E_{ad} = \gamma_{gs} S = \gamma_{gs} \pi R^{2}.
\label{eq:adhesion}
\end{equation}
\noindent
where $\gamma_{gs}$ is the specific adhesion energy between the graphene sheet and the substrate, $S$ is the footprint area and $R$ is the GNB footprint radius.

The total energy $E_{total}$ of the system consists of three parts: the energy of the confined fluid, the elastic energy of the graphene sheet and the adhesion energy:

\begin{equation}
    E_{total} = E_{cf} + E_{el} + E_{ad}.
\end{equation}

To obtain an equilibrium nanobubble with a certain mass of the confined fluid, the total energy is minimized according to "the shape" of the GNB and the density profile of the confined fluid $\rho(z)$. "The shape" concept of the GNB includes the footprint radius, the height profile $h(r)$ and the displacement profile $u(r)$. In addition to the mass constraint imposed on the total energy minimization, there is an extra condition of mechanical equilibrium between the graphene sheet and the confined fluid, which implies that the pressure applied to the confined fluid is equal to the pressure developed by the graphene membrane (Fig.~\ref{fig:fig1}b).

The elastic energy of the graphene sheet is defined in terms of elasticity theory~\cite{landau1959course}. The energy expression from the theory of equilibrium of plates applied in cylindrical coordinates yields the following expression:

\begin{equation}
    E_{el} = \int\limits_{0}^{\infty} \left[\Psi_{s} (u_{\alpha \beta}, h) + \Psi_{b} (h) \right] \, 2 \pi r dr,
\end{equation}

\noindent where $ \Psi_{s} $ and $ \Psi_{b} $ are the stretching and bending energies per unit area, respectively; $ u_{\alpha \beta} $ is the strain tensor; $ h $ is the height profile; and $ r $ is radius. Additionally, from the theory of elasticity, the pressure developed by the graphene sheet deformation is calculated as follows:

\begin{equation}
    P_{mech} = \frac{\sigma_{rr}}{r_r} + \frac{\sigma_{\theta \theta}}{r_{\theta}},
    \label{eq:pressure_mech}
\end{equation}

\noindent where $r_r$, $r_{\theta}$ represent the principal radii of the curvature of the graphene membrane at a particular point and $\sigma_{rr}$, $\sigma_{\theta\theta}$ are the radial and angular stress distributions in the graphene membrane, respectively.

To evaluate the internal energy of the confined fluid by means of classical DFT, the bubble profile is reduced to the slit geometry representation (Fig.~\ref{fig:fig1}a). The height of the slit geometry is evaluated as:

\begin{equation}
    H_{slit} = V/\pi R^2, 
    \label{eq:h_slit}
\end{equation}

\noindent
where $V$ is the volume of the GNB. Fundamental measure theory~\cite{roth2002fundamental}, which is a version of classical DFT, is employed to calculate the specific Helmholtz free energy $f$ and the density profile $\rho(z)$ in the direction normal to the surface. Then, the energy of the confined fluid is simply given by:
\begin{equation}
    E_{cf} = f(H_{slit}, \mu, T) V.
    \label{eq:cf_energy}
\end{equation}
\noindent where $\mu$ is the bulk chemical potential and $T$ is the temperature. The pressure $P_{cf}$ inside the confined fluid is calculated according to the procedure described in the work~\cite{gregoire2018estimation}:
\begin{equation}
    P_{cf} = -\frac{1}{\pi R^2}\frac{\partial \Omega}{\partial H_{slit}},
    \label{eq:solvation_pressure}
\end{equation}
where $\Omega[\rho]$ is the grand canonical potential that corresponds to the equilibrium confined fluid state $\rho(z)$.

\begin{figure}
\includegraphics[width=0.9\columnwidth]{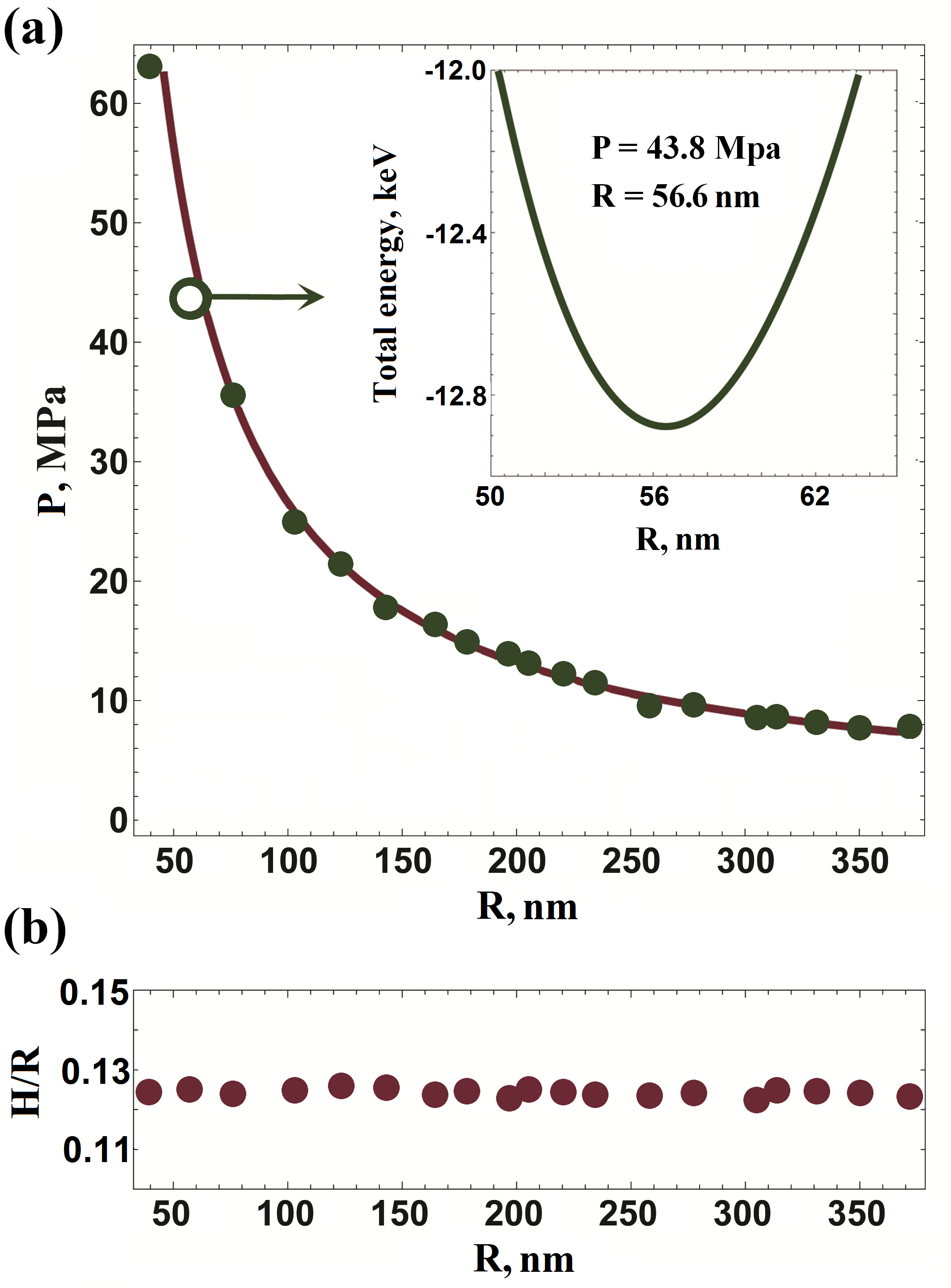}
\caption{(a) Line of equilibrium GNBs filled with argon in ($P$, $R$) coordinates. Calculations are performed for masses that correspond to a number of Ar atoms ranging from $0.2 \times 10^6$ to $24 \times 10^6$. Inset graph: characteristic example of the total energy profile for the isomass curve of a GNB storing $0.5 \times 10^6$ atoms. (b) Calculated height-to-radius ratio $H/R$. It is almost constant in the considered range of radii and equal to 0.125.}
\label{fig:fig2}
\end{figure}

The algorithm of the total energy evaluation of a GNB consists of two steps (see Fig.~\ref{fig:fig1}b). In the first step, extensive tabulated data are generated for the elastic, confined fluid and adhesion parts of the total energy. Input parameters for every contribution are taken in a specified range, and corresponding meshes are built. The mesh of the elastic energy contribution is built in ($P_{mech}, R)$ coordinates, the mesh of the confined fluid contribution is generated in $(P_{cf}, H_{slit})$ coordinates, and the 1D mesh of the adhesion contribution is constructed in an ($R$) mesh. After the meshes are generated for every point, the output parameters of each energy contribution are calculated.

In the second step of the algorithm, the points of the obtained meshes are joined together. The procedure is as follows. A particular point in $(P_{mech}, R)$ is chosen because the calculation of $E_{el}$ is already performed. The volume $V$ of the system is also known. Then, the point in $(P_{cf}, H_{slit})$ that meets the following two conditions is found: $P_{mech} = P_{cf}$ and $H_{slit} = V / \pi R^2 $. At this stage, the density profile $\rho(z)$ is obtained from classical DFT, and the average density of the confined fluid is evaluated in the $z$ direction:
\begin{equation}
\label{eq:dens_cf}
    \rho_{cf} = \frac{1}{H_\text{slit}}\int \rho(z) \; dz.
\end{equation}
\noindent
The mass $M$ of the bubble is simply evaluated as $\rho_{cf} V$. The joining with the remaining 1D $R$-mesh of adhesion is performed by choosing the point with the corresponding radius. 

The procedure described above leads to numerous GNBs with the same mass but different radii, internal pressures and total energies. They can be depicted as isomass curves in $(P, R)$ coordinates (see Fig.\ref{fig:fig1}c and Fig.\ref{fig:fig2}b). The equilibrium GNB on the isomass curve is determined as the bubble with the lowest total energy.

\begin{figure}
\includegraphics[width=0.92\columnwidth]{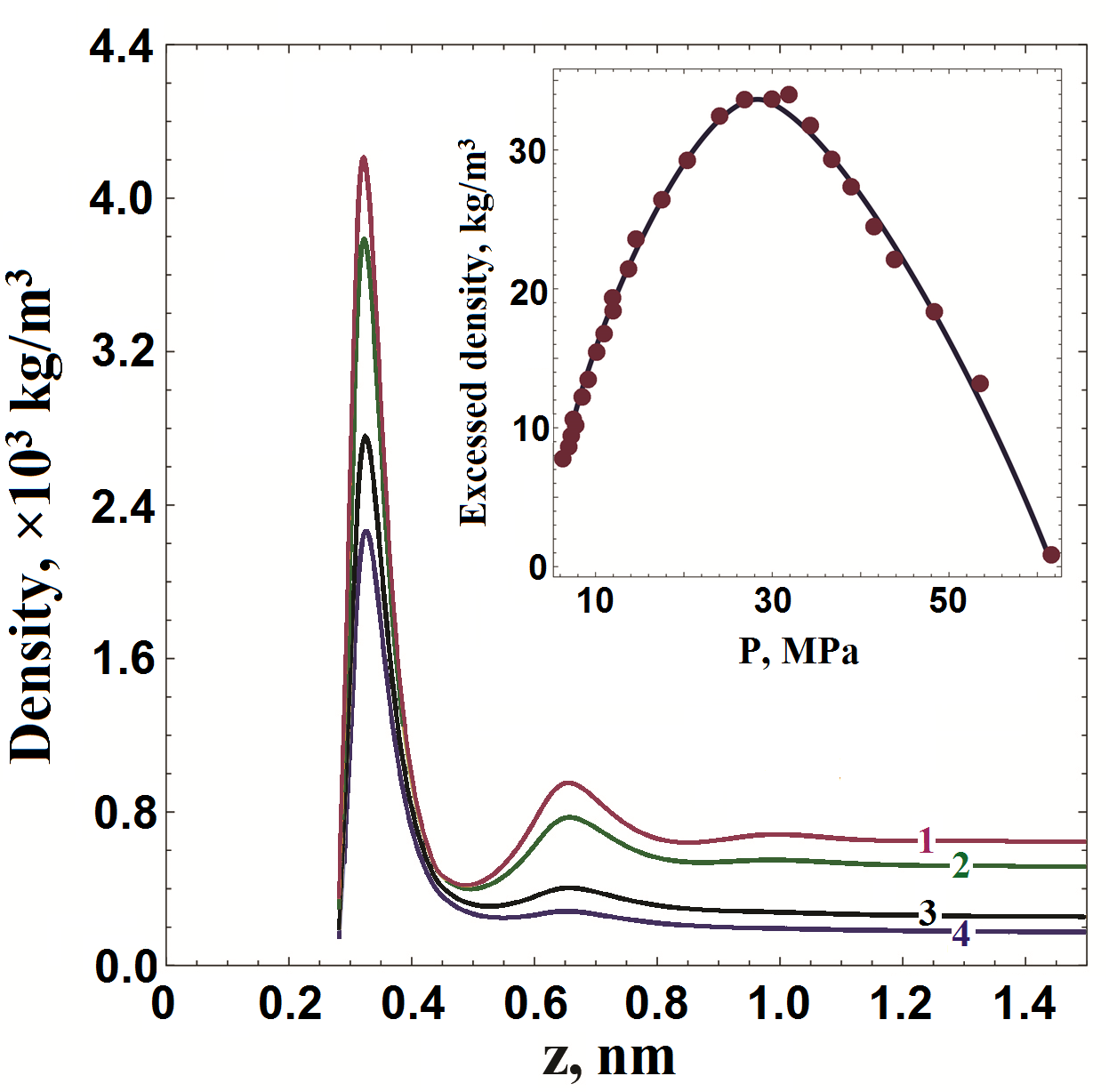}
\caption{\label{fig:fig3} Density distribution along the z axis obtained from classical DFT. Curves from 1 to 4 correspond to the following slit geometry parameters ($H_{slit}$, $P$):  (4.38~nm, 43.79 MPa), (6.7~nm, 31.74~MPa), (14.4~nm, 14.57~MPa) and (20.5~nm, 10.17~MPa), respectively. Inset graph: excess density at different pressures.}
\end{figure}

The developed model is applied to GNBs on a graphite substrate with trapped argon atoms. The temperature is fixed in all calculations and equals $300~K$. The resulting equilibrium GNBs in $(P, R)$ coordinates are presented in Fig.~\ref{fig:fig1}c and with additional details in Fig.~\ref{fig:fig2}a. In accordance with our algorithm, the equilibrium GNB corresponds to the minimum of the isomass total energy, as shown in the inset graph in Fig.~\ref{fig:fig2}a. GNBs in the range of 50--350~nm correspond to pressures in the range of 60--8~MPa. A significant increase in the pressure is observed for small GNBs ($R <$ 100~nm), which is in accordance with previous multiscale modeling~\cite{zhilyaev2019liquid} and MD simulations~\cite{iakovlev2017atomistic}.

The standard approach to characterize GNB profiles in AFM experiments is to measure the height-to-radius ($H/R$) ratio. One study~\cite{khestanova2016universal} shows that the $H/R$ ratio is universal, i.e., independent of the bubbles’ radius or volume. Our algorithm provides a geometric profile of the equilibrium GNBs that also allows calculation of the height and radius. The constant ratio $ H/R = 0.125 $ is observed for all considered equilibrium GNBs (see Fig.~\ref{fig:fig2}b) from our calculations. This result is in agreement with experimental data~\cite{khestanova2016universal} and previous theoretical studies~\cite{iakovlev2017atomistic, zhilyaev2019liquid}.

The developed model describes both the external equilibrium GNB properties (P, R) and the corresponding inner structure. The confined fluid structure is characterized by the density distribution obtained from the DFT part of the model. The classical DFT approach calculations allow us to evaluate $\rho_{cf}(z)$ for a given height of the slit geometry and pressure. The typical density distributions are shown in Fig.~\ref{fig:fig3} for various GNBs. Confined liquid structures can also be characterized in terms of excess density:

\begin{equation}
    \rho_{exc} = \rho_{cf} - \rho_{0},
\end{equation}

\noindent where $\rho_{0}$ is the bulk fluid density for the corresponding pressure. In Fig.~\ref{fig:fig3}, the inset graph shows the nonmonotonic behavior of $\rho_{exc}$ for different equilibrium GNB pressures. All considered GNBs exhibit a positive excess density that corresponds to the presence of a dense layer near the surface. Nevertheless, if one extrapolates the excess density graph in the region of high pressures, which corresponds to GNBs with radii less than 50~nm, the excess density becomes negative. We assume that negative excess density could be the precursor to the liquid-solid phase transition. This prediction agrees well with recent molecular dynamics simulations~\cite{iakovlev2017atomistic} in which solid argon at room temperature was observed in GNBs with radii less than 35~nm.

One of the limitations of the developed model is that it cannot consider crystal structures that presumably arise for GNBs with radii less than 50~nm. There are two reasons for this limitation. First, an assumption about the isotropic pressure that is produced by the confined fluid is built into the model. Once the trapped substance is solid, this is not the case, and one has to consider the anisotropic pressure tensor. As a result, the elastic problem for the graphene membrane became significantly harder to solve. Second, classic DFT itself is still a limited method for considering both confined liquid and solid phases simultaneously. The currently available liquid/solid-state DFT frameworks are still in the beginning stage of development~\cite{kocher2015new}.

Nevertheless, the case of the solidification of the substance trapped inside GNBs is quite interesting. It is known both experimentally~\cite{zamborlini2015nanobubbles, larciprete2016self} and from MD simulations~\cite{iakovlev2017atomistic} that solid structures could emerge inside small GNBs with radii of less than approximately 50~nm. Additionally, small GNBs could exist in different morphological forms, for example, experimentally and MD observed "pancake" or "flat island" forms~\cite{khestanova2018van, iakovlev2017atomistic}. Thus, the development of a comprehensive GNB model that should be able to describe gas-, liquid- and solid-phase states of a trapped substance simultaneously is in demand and would be the objective of future research.

In conclusion, we developed a model of GNBs that takes into account the inhomogeneous structure of the confined fluid and the mechanical stretching of the graphene membrane. The classical DFT approach is used to describe confined fluid properties. It provides information on the inner structure in terms of density profiles. The graphene sheet mechanics with a constant pressure profile are described by the theory of elasticity of membranes. Unlike the confined fluid theory in a solid porous medium, a GNB is an example of nonrigid confinement that requires an additional condition of mechanical equilibrium. We develop an algorithm that calculates GNB characteristics by taking into account the equilibrium of the graphene membrane and trapped fluid in a self-consistent manner. As an example application, we consider GNBs filled with argon at room temperature in the radii range of 50--350~nm. Our calculations present the universal shape of equilibrium GNBs for the whole range of radii, i.e., constant $H/R$ ratio, which is consistent with previous experimental measurements and MD simulations. Additionally, the density profiles and excess density of the GNBs in the considered range of radii are evaluated. It is shown that the extrapolation of the obtained results to the region of radii smaller than 50~nm leads to negative excess densities, which can presumably be attributed to the onset of the liquid-solid phase transition.

\begin{acknowledgments}
The research was partially carried out using supercomputers at the Joint Supercomputer Center of the Russian Academy of Sciences (JSCC RAS). The Authors also acknowledge the usage of the Skoltech CDISE HPC cluster "ARKUDA" for partly obtaining the results presented in this paper.
\end{acknowledgments}

\appendix

\nocite{*}
\bibliography{aipsamp}

\end{document}